\documentclass[twocolumn, aps, pra,prl, superscriptaddress, preprintnumbers, letterpaper]{revtex4-1}

\usepackage{amsthm}
\usepackage{amsmath,bm}
\usepackage{amssymb}
\usepackage{braket}
\usepackage{bbold}
\usepackage{graphicx}
\usepackage{subfigure}
\usepackage{hyperref}
\usepackage{appendix}
\usepackage{tikz}
\usepgflibrary{shapes.geometric}
\usetikzlibrary{arrows}

\hypersetup{colorlinks=true, citecolor=blue, urlcolor=blue, linkcolor=blue}

\newcommand{\norm}[1]{\lVert#1\rVert}
\begin{document}
\title{Genuine Multipartite Entanglement Measure Based on $\alpha$-concurrence}
\author{Ke-Ke Wang}
\email{wangkk@cnu.edu.cn}
\affiliation{School of Mathematical Sciences, Capital Normal University, 100048 Beijing, China}
\author{Zhi-Wei Wei}
\email{weizhw@cnu.edu.cn}
\affiliation{School of Mathematical Sciences, Capital Normal University, 100048 Beijing, China}
\author{Shao-Ming Fei}
\email{feishm@cnu.edu.cn (Corresponding author)}
\affiliation{School of Mathematical Sciences, Capital Normal University, 100048 Beijing, China}

\bigskip

\begin{abstract}
Quantifying genuine entanglement is a crucial task in quantum information theory.
Based on the geometric mean of bipartite $\alpha$-concurrences among all bipartitions, we present a class of well-defined genuine multipartite entanglement (GME) measures G$\alpha$C with one parameter $\alpha$ for arbitrary multipartite states. We show that the G$\alpha$C is of continuity for any multipartite pure states. By utilizing the related symmetry, analytical results of G$\alpha$C are derived for any $n$-qubit GHZ states and W states, which show that the GHZ states are more genuinely entangled than the W states. With explicit examples, we demonstrate that the G$\alpha$C can distinguish different GME states that other GME measures fail to. These results justify the potential applications of G$\alpha$C in characterizing genuine multipartite entanglements.
\end{abstract}

\maketitle

\section{introduction}
{\color{blue}
The quantum resource theory \cite{RMP91025001} contains three issues, the free states, resource states and the free operations which transform free states into free states. A measure of resource is a real function of the quantum states, which is greater zero for any resource states and zero for all free states. Moreover, a measure of resource does not increase under free operations. Quantum entanglement is a specific and critically important resource in many quantum information processing protocols and quantum computation \cite{nielsen2002quantum,RMP81865}. In the resource theory of quantum entanglement, the free states are the separable states, while the resource states are the entangled ones. The free operations are given by the local operations and classical communication (LOCC). The entanglement measures are just the measures of the resource of quantum entanglement \cite{vidal2000entanglement,ma2011measure,Revisevidal2000ent}. The quantum entanglement is the essential resources in quantum information processing such as quantum cryptography \cite{PRL67661,PRL68557,PRA572383,PRL102140501}, quantum teleportation \cite{PRL701895,epjb4175,Gao2008}, quantum dense coding  \cite{PRL692881}, clock synchronization \cite{PRL852010} and quantum communication \cite{hiesmayr2009two,shimoni2005entangled}. Therefore, the quantification of entanglement in bipartite and multipartite systems is a key task in the theory of quantum entanglement.}

Many entanglement measures for bipartite systems have been proposed \cite{Vidal,zhixiang1,guhne2005multipartite}, such as concurrence \cite{PhysRevLett.78.5022,PhysRevA.64.042315}, entanglement of formation \cite{PhysRevA.54.3824,PhysRevLett.80.2245,horodecki2001entanglement},
negativity \cite{PhysRevA.65.032314,PhysRevLett.95.090503}, R\'enyi-$\alpha$ entropy of entanglement \cite{gour2007dual,Kim_2010} and Tsallis entropy of entanglement \cite{raggio1995properties,PRA81062328}. In particular, recently two types of parameterized bipartite entanglement measures $q$-concurrence ($q\geqslant2$) \cite{yang2021parametrized} and $\alpha$-concurrence ($0\leqslant\alpha\leqslant1/2$) \cite{Wei} have been presented based on Tsallis entropy \cite{Tsallise,PLA247211}.

The situation for multipartite systems becomes more complicated with the increase of the number of parties and the dimension of the subsystems. A special kind of multipartite entanglement is the genuine multipartite entanglement. Many efforts have been made toward the quantification of genuine multipartite entanglement (GME) in the last few decades  \cite{Coffman,Barnum,blasone2008hierarchies,Sen,ma2011measure,xie2021triangle}.
Based on the shortest distance from a given state to the $k$-separable states, a GME measure is proposed as the generalized geometric measure (GGM) \cite{blasone2008hierarchies,Sen}.
The genuine multipartite concurrence (GMC) is defined as the minimal bipartite concurrence among all bipartitions \cite{ma2011measure}. The GME measure of three-qubit states is introduced in \cite{xie2021triangle} based on the area of the three-qubit concurrence triangle. In \cite{Guoyu22}, the authors generalized the above method to multipartite case.
Recently, the geometric mean of bipartite concurrence (GBC) \cite{li2022geometric} is used as a GME measure, which is given by the geometric mean of bipartite concurrence among all bipartitions, which is further generalized to parameterized form, known as the geometric mean of $q$-concurrence (G$q$C) \cite{Xian}.

Inspired by the approaches in \cite{li2022geometric,Xian}, we aim to introduce a new concept G$\alpha$C based on the geometric mean of $\alpha$-concurrence ($0\leqslant\alpha\leqslant1/2$). We show that the G$\alpha$C is a class of well-defined GME measures with one parameter $\alpha$ for multipartite states, which gives rise to a better characterization of multipartite entanglement. We further show that the G$\alpha$C is qualified for quantifying potential quantum resources, benefiting from desirable properties such as continuity. By utilizing the symmetry in the definition of the G$\alpha$C, analytical results are obtained for any $n$-qubit GHZ states and W states. By detailed examples, we demonstrate that the G$\alpha$C can distinguish the difference in certain features of GME while other GME measures cannot.

The paper is structured as follows. In Sec.~\ref{se2}, we briefly introduce some basic concepts and notions. In Sec. \ref{se2-1}, we introduce a new concept G$\alpha$C, which is a class of well-defined GME measures based on the geometric mean of $\alpha$-concurrence.
We further show that the G$\alpha$C is of continuity for any multipartite pure states.
Analytical results are obtained for any $n$-qubit GHZ states and W states.
In Sec. \ref{s3} we study the ordering of G$\alpha$C compared with some other GME measures.
We summarize and discuss in Sec. \ref{se4}.

\section{Preliminary}\label{se2}
Let $\mathcal{H}_{A_i}$ denote the $d$-dimensional Hilbert space associated with the $i$th subspace of an $n$-partite system. An $n$-partite pure state $\ket{\psi}\in\mathcal{H}_{A_1}\otimes\mathcal{H}_{A_2}\otimes\cdots\otimes\mathcal{H}_{A_n}$ is called bi-separable if it can be written as $\ket{\psi}=\ket{\psi_{S}}\otimes\ket{\psi_{\bar{S}}}$, where $\ket{\psi_{S}}\in\mathcal{H}_{S}=\mathcal{H}_{A_{j_1}}\otimes\cdots\otimes\mathcal{H}_{A_{j_k}}$ and $\ket{\psi_{\bar{S}}}\in\mathcal{H}_{\bar{S}}=\mathcal{H}_{A_{j_{k+1}}}\otimes\cdots\otimes\mathcal{H}_{A_{j_n}}$ under some bipartition $S|\bar{S}$. An $n$-partite mixed state $\rho$ is biseparable if it can be written as a convex combination of biseparable pure states $\rho=\sum_i p_i|\psi_i\rangle\langle\psi_i|$, wherein $\{\ket{\psi_i}\}$ may be biseparable with respect to different bipartitions. If an $n$-partite state is not biseparable, then it is called genuine $n$-partite entangled. We denote below the set of all multipartite states by $\mathcal{D}$.

A well-defined GME measure $E(\rho)$ of state $\rho$ should satisfy following conditions \cite{vidal2000entanglement,ma2011measure,Revisevidal2000ent}:
\begin{itemize}
\item[(M1)] $E(\rho)\geqslant0$ for any state $\rho\in\mathcal{D}$, where the equality holds only for biseparable states.
\item[(M2)] $E$ is convex, $E(\sum_ip_i\rho_i)\leqslant\sum_ip_iE(\rho_i)$.
\item[(M3)] Monotonicity, $E$ cannot increase under LOCC, $E(\rho)\geqslant E(\Lambda(\rho))$ for any LOCC map $\Lambda$.
\item[(M4)] $E$ is invariant under local unitary transformations, $E(\rho)=E(U_{local}\rho U_{local}^{\dagger})$ for any local unitaries $U_{local}$.
\item[(M5)] Strong monotonicity, $E$ cannot increase on average under LOCC operations,
\begin{equation}
E(\rho)\geqslant\sum_ip_iE(\sigma_i),
\end{equation}
where $\{p_i,\sigma_i\}$ is an ensemble generated by an arbitrary LOCC operation $\Phi_\text{LOCC}$,
\begin{equation}
\Phi_\text{LOCC}(\rho)=\sum_ip_i\sigma_i.
\end{equation}
\end{itemize}

The $\alpha$-concurrence of a bipartite pure state $\ket{\psi}\in\mathcal{H}_{A_1}\otimes\mathcal{H}_{A_2}$ is given by \cite{Wei}
\begin{equation}\label{text11}
C_{\alpha}\left(\ket{\psi}\right)=\mathrm{Tr}\rho_{A_1}^{\alpha}-1
\end{equation}
for any $0\leqslant \alpha\leqslant1/2$, where $\rho_{A_1}=\mathrm{Tr}_{A_2}|\psi\rangle\langle\psi|$.
For a general mixed state $\rho\in\mathcal{H}_{A_1}\otimes\mathcal{H}_{A_2}$, the $\alpha$-concurrence is given by the convex-roof extension,
\begin{equation}\label{text14}
C_{\alpha}(\rho)=\min_{\left\{p_i,\ket{\psi_i}\right\}}
\sum_ip_iC_{\alpha}\left(\ket{\psi_i}\right),
\end{equation}
where the infimum is taken over all possible pure state decompositions of $\rho=\sum_ip_i|\psi_i\rangle\langle\psi_i|$, with $\sum_ip_i=1$ and $p_i> 0$.

\section{Geometric mean of $\alpha$-concurrence}\label{se2-1}
We now introduce a measure of multipartite entanglement satisfying all necessary conditions ((M1)-(M5)) based on the $\alpha$-concurrence. Let $\ket{\psi}\in\otimes_{k=1}^n\mathcal{H}_{A_k}$ be an arbitrary $n$-partite pure state. We define the geometric mean of $\alpha$-concurrence (G$\alpha$C) by
\begin{align}
\mathcal{G}_\alpha(\ket{\psi})=\sqrt[{c(\beta)}]{\mathcal{P}_\alpha(\ket{\psi})},
\end{align}
where $\beta=\{\beta_j\}$ denotes the set of all possible bipartitions $\{S_{\beta_j}|\bar{S}_{\beta_j}\}$ of the $n$-parties, $c(\beta)$ is the cardinality of $\beta$, and $\mathcal{P}_\alpha(\ket{\psi})$ is the product of all bipartite $\alpha$-concurrences,
$$
\mathcal{P}_\alpha(\ket{\psi})=\Pi_{\beta_j\in\beta}
\mathcal{C}_\alpha^{S_{\beta_j}|\bar{S}_{\beta_j}}(\ket{\psi}),
$$
\begin{align*}
	\begin{split}
	c(\beta)=\left\{
	\begin{array}{lr}
\sum_{m=1}^{\frac{n-1}{2}}C_n^m,\hspace{3mm} \text{if}~ n ~\text{is}~ \text{odd},\\[1mm]
\sum_{m=1}^{\frac{n-2}{2}}C_n^m+\frac{1}{2}C_n^{\frac{n}{2}},\hspace{3mm}\text{if}~ n ~\text{is}~\text{even}.
\end{array}
\right.
\end{split}
\end{align*}
For a general mixed state $\rho\in \otimes_{k=1}^n\mathcal{H}_{A_k}$, the G$\alpha$C is given by the convex-roof extension,
\begin{align}\label{ga1}
\mathcal{G}_\alpha(\rho)=\min\sum_i p_i\mathcal{G}_\alpha(\ket{\psi_i}),
\end{align}
where the infimum is taken over all possible pure state decompositions of $\rho=\sum_ip_i|\psi_i\rangle\langle\psi_i|$, with $\sum_ip_i=1$ and $p_i> 0$.

$\mathit{Theorem\ 1}$.
The geometric mean of $\alpha$-concurrence $\mathcal{G}_\alpha(\rho)$ given in (\ref{ga1}) is a well defined genuine multipartite entanglement measure.

$\mathit{Proof}$.
We need to verify that $\mathcal{G}_\alpha(\rho)$ fulfills the following requirements.

(M1) The definition of $\mathcal{G}_\alpha(\rho)$ directly implies that $\mathcal{G}_\alpha(\rho)=0$ for all biseparable states and $\mathcal{G}_\alpha(\rho)>0$ for all genuine $n$-partite entangled states.

(M2) The convexity  follows directly from the fact that any mixture $\lambda\rho_1+(1-\lambda)\rho_2$ of two density matrices $\rho_1$ and $\rho_2$ is at least decomposable into states that attain the individual infima.

(M3) As the $\alpha$-concurrence of any subsystem has been proven to be nonincreasing under LOCC \cite{Wei}, the minimum of all possible $\alpha$-concurrences remains nonincreasing too, thus proving that $\mathcal{G}_\alpha(\rho)\geqslant\mathcal{G}_\alpha(\Lambda(\rho))$ for any LOCC map $\Lambda$.

(M4) Consider $\rho'=U_{local}\rho U_{local}^{\dagger}$. From condition (M3) we have $\mathcal{G}_\alpha(\rho)\geqslant\mathcal{G}_\alpha(\rho')
\geqslant\mathcal{G}_\alpha(U_{local}^{\dagger}\rho'U_{local})=\mathcal{G}_\alpha(\rho)$. Hence, $\mathcal{G}_\alpha(\rho)$ is invariant under local unitary transformations.

(M5) The bipartite $\alpha$-concurrence satisfies the property of average monotonicity \cite{Wei},
\begin{equation}\label{m5-1}
C_{\alpha}^{S|\bar{S}}(\rho)\geqslant\sum_ip_iC_{\alpha}^{S|\bar{S}}(\sigma_i),
\end{equation}
where $\{p_i,\sigma_i\}$ is an ensemble generated by an arbitrary LOCC channel $\Phi_\text{LOCC}$ acting on the multipartite quantum state $\rho$, such that $\Phi_\text{LOCC}(\rho)=\sum_ip_i\sigma_i$.
Note that any LOCC is also an LOCC channel with respect to any bipartition $\{S|\bar{S}\}$. First, we examine the average monotonicity for pure states. Since all the $\alpha$-concurrence in the definition of G$\alpha$C satisfy (\ref{m5-1}), we have
\begin{align}\label{a3}
\mathcal{G}_\alpha(\rho)&=\sqrt[{c(\beta)}]{\mathcal{P}_\alpha(\rho)}\nonumber\\
&=\sqrt[{c(\beta)}]{\Pi_{\beta_j\in\beta}C_\alpha^{S_{\beta_j}|\bar{S}_{\beta_j}}(\rho)}\nonumber\\
&\geqslant\sqrt[{c(\beta)}]{\Pi_{\beta_j\in\beta}(\sum_{i} p_{i}C_\alpha^{S_{\beta_j}|\bar{S}_{\beta_j}}(\sigma_i))}\nonumber\\
&\geqslant\sum_{i} p_{i}\sqrt[{c(\beta)}]{\Pi_{\beta_j\in\beta}C_\alpha^{S_{\beta_j}|\bar{S}_{\beta_j}}(\sigma_i))}\nonumber\\
&=\sum_{i} p_{i}\mathcal{G}_{\alpha}(\sigma_i),
\end{align}
where the second inequality is duet to the Mahler's inequality. The average monotonicity of mixed states is naturally inherited from the monotonicity of pure states via the convex roof extension \cite{vidal2000entanglement}.
$\hfill\qedsymbol$

The G$\alpha$C is a class of GME measures with parameter $\alpha$. Next, we prove that the G$\alpha$C is continuous for multipartite pure states. We first prove that the $\alpha$-concurrence is continuous for bipartite pure states.

$\mathit{Lemma\ 1.}$
Let $\ket{\psi_1}$ and $\ket{\psi_2}$ be two bipartite pure states in $\mathcal{H}_{A_1}\otimes\mathcal{H}_{A_2}$ with $\|\ket{\psi_1}-\ket{\psi_2}\|_1\leqslant \epsilon$, where $\|X\|_1$ is the trace-norm of $X$ defined by $\|X\|_1=\text{Tr}\sqrt{XX^{\dagger}}$. We have
\begin{align}\label{l0}
|\mathcal{C}_\alpha(\ket{\psi_1})-\mathcal{C}_\alpha(\ket{\psi_2})|\leqslant\epsilon^\alpha d^{1-\alpha}.
\end{align}

$\mathit{Proof}$. Since partial trace is trace-preserving, we have $\|\rho_{A_1}-\sigma_{A_1}\|_1\le \epsilon$, where $\rho_{A_1}=\text{Tr}_{A_2}\ket{\psi_1}\bra{\psi_1}$ and $\sigma_{A_1}=\text{Tr}_{A_2}\ket{\psi_2}\bra{\psi_2}$.
As the trace-norm is a unitarily invariant norm \cite{bhatia2013matrix}, we have
\begin{align}
\norm{\text{Eig}^{\downarrow}(\rho_{A_1})-\text{Eig}^{\downarrow}(\sigma_{A_1})}_1\le \norm{\rho_{A_1}-\sigma_{A_1}}_1,\nonumber\\
\norm{\rho_{A_1}-\sigma_{A_1}}_1 \le \norm{\text{Eig}^{\downarrow}(\rho_{A_1})-\text{Eig}^{\uparrow}(\sigma_{A_1})}_1,
\end{align}
where $\text{Eig}^{\downarrow}(X)$ ($\text{Eig}^{\uparrow}(X)$) denotes
the diagonal matrix given by the eigenvalues $\{x_1,\cdots,x_d\}$ of $X$ such that $\{|x_1|\geq\cdots\geq |x_d|\}$ ($\{|x_1|\leq\cdots\leq |x_d|\}$).

Let $\{p_i\}_{i=1}^d$ and $\{r_i\}_{i=1}^d$ be the eigenvalues of $\rho_{A_1}$ and $\sigma_{A_1}$, respectively, such that $\sum_ip_i=\sum_ir_i=1$ and $\sum_i|p_i-r_i|\leq\epsilon$. Then
\begin{align}
|\mathcal{C}_\alpha(\ket{\psi_1})-\mathcal{C}_\alpha(\ket{\psi_2})|=&|\mathrm{Tr}\rho_{A_1}^\alpha-\mathrm{Tr}\sigma_{A_1}^\alpha|\nonumber\\
=&|\sum_i(p_i^\alpha-r^\alpha_i)|\nonumber\\
\leq &\sum_i|p_i^\alpha-r^\alpha_i|.\label{in1}
\end{align}
Set $|p_i-r_i|=\epsilon_i$. Then $\sum_{i=1}^d \epsilon_i\leq \epsilon$. When $p_i=r_i+\epsilon_i$, we have
	$$
	|p_i^\alpha-r_i^\alpha|=|(r_i+\epsilon_i)^\alpha-r_i^\alpha|\leq (\epsilon_i)^\alpha,
	$$
where the last inequality is due to that $(r_i+\epsilon_i)^\alpha-r_i^\alpha$ is decreasing order withe respect to $r_i$, and $r_i\in(0,1).$  Then the inequality $(\ref{in1})$ becomes
$$
|\mathcal{C}_\alpha(\ket{\psi_1})-\mathcal{C}_\alpha(\ket{\psi_2})
|\leq\sum_i|p_i^\alpha-r^\alpha_i|
	\leq \epsilon^\alpha d^{1-\alpha}.
$$
The last equality holds when all $\epsilon_i=\frac{\epsilon}{d}$.
$\hfill\qedsymbol$

Generalizing the results to the multipartite case, we have

$\mathit{Theorem\ 2}$.
Let $\ket{\psi_1}$ and $\ket{\psi_2}$ be two multipartite pure states in $\otimes_{k=1}^n\mathcal{H}_{A_k}$. If $\norm{\ket{\psi_1}-\ket{\psi_2}}_1\le \epsilon$, then
\begin{align}
	|\mathcal{G}_\alpha(\ket{\psi_1})-\mathcal{G}_\alpha(\ket{\psi_2})|\le [\sum_{i=1}^{\frac{n-1}{2}}C_n^i\epsilon^\alpha d_i^{1-\alpha}]^{\frac{1}{c(\beta)}},
\end{align}
where $d_i$ represents the minimum dimension of the two subsystems in the $i$-th bipartition.

$\mathit{Proof}$.
When $n$ is odd, we have
\begin{align}
&|\mathcal{G}_\alpha(\ket{\psi_1})-\mathcal{G}_\alpha(\ket{\psi_2})|\nonumber\\
=&|\mathcal{P}_\alpha(\ket{\psi_1})^{\frac{1}{c(\beta)}}-\mathcal{P}_\alpha(\ket{\psi_2})^{\frac{1}{c(\beta)}}|\nonumber\\
\le&|\mathcal{P}_\alpha(\ket{\psi_1})-\mathcal{P}_\alpha(\ket{\psi_2})|^{\frac{1}{c(\beta)}}\nonumber\\
=&|\Pi_{\beta_j\in\beta}\mathcal{C}_\alpha^{S_{\beta_j}|\bar{S}_{\beta_j}}(\ket{\psi_1})-\Pi_{\beta_j\in\beta}\mathcal{C}_\alpha^{S_{\beta_j}|\bar{S}_{\beta_j}}(\ket{\psi_2})|^{\frac{1}{c(\beta)}}\nonumber\\
\le&[\sum_{i=1}^{\frac{n-1}{2}}C_n^i\epsilon^\alpha d_i^{1-\alpha}]^{\frac{1}{c(\beta)}},
\end{align}
where the first inequality is due to the fact that $|p^x-q^{x}|\le |p-q|^x$ when $p,q,x\in(0,1)$,
the second inequality is due fact
\begin{align}
&|x_1x_2\cdots x_k-y_1y_2\cdots y_k|\nonumber\\
=& |(x_1-y_1)x_2x_3\cdots x_k+y_1(x_2-y_2)x_3\cdots x_k\nonumber\\
&+y_1y_2(x_3-y_3)x_4\cdots x_k\cdots+y_1y_2\cdots y_{k-1}(x_k-y_k)|\nonumber\\
\le& \sum_i|x_i-y_i|
\end{align}
for $x_i,y_i\in(0,1)$, $i=1,2,\cdots,k$, the last inequality is due to the triangle inequality for $x_i,y_i\in(0,1)$.
$\hfill\qedsymbol$

A pure multipartite entangled state is called an absolutely maximally entangled (AME) state if all reduced density operators obtained by tracing over half of the subsystems are proportional to the identity \cite{PRA86052335}. The AME states are of significance in many quantum information processing tasks such as quantum secret sharing schemes \cite{PRA86052335} and quantum error correction codes \cite{PRA69052330}. Specifically, the three-qubit AME state, i.e., the GHZ state, gives rise to perfect quantum teleportation \cite{goyeneche2015absolutely,shi2021multilinear}, while the W state achieves only a maximal success rate of 2/3 \cite{joo2003quantum}. Here, the G$\alpha$C for each parameter $\alpha$ takes the maximal values for all the AME states. We consider in the following example the $n$-partite ($n\ge 3$) GHZ and W states that are not equivalent under stochastic local operations and classical communication (SLOCC).

$\mathit{Example\ 1}$.
Consider the $n$-partite W state and GHZ state,
\begin{align*}
&\ket{W_n}=\frac{1}{\sqrt{n}}(\ket{10\cdots0}+\ket{01\cdots0}+\cdots+\ket{00\cdots1}),\\
&\ket{GHZ_n}=\frac{1}{\sqrt{2}}(\ket{00\cdots0}+\ket{11\cdots1}).
\end{align*}
The $\alpha$-concurrence under the bipartition of the $k$th qubit and the rest is given by
$$
\mathcal{C}_{\alpha,k,n-k}(\ket{W_n})=\frac{k^\alpha}{n^\alpha}+\frac{(n-k)^\alpha}{n^\alpha}-1
$$
for the state $\ket{W_n}$, and $2^{1-\alpha}-1$ for the state $\ket{GHZ_n}$.
Therefore, we have $\mathcal{G}_\alpha\ket{GHZ_n})=2^{1-\alpha}-1$ and
\begin{widetext}
$$
\begin{array}{rcl}
	\mathcal{G}_\alpha(\ket{W_n})&=&
	\begin{split}
	\left\{
	\begin{array}{lr}
 (\Pi_{k=1}^{\frac{n-1}{2}}[\frac{k^\alpha}{n^\alpha}+\frac{(n-k)^\alpha}{n^\alpha}-1]^{C_n^k})^{\frac{1}{c(\beta)}},\hspace{3mm} \text{if $n$ is odd}\\ (\Pi_{k=1}^{\frac{n}{2}-1}[\frac{k^\alpha}{n^\alpha}+\frac{(n-k)^\alpha}{n^\alpha}-1]^{C_n^k}[2^{1-\alpha}-1]^{\frac{C_n^{\frac{n}{2}}}{2}})^{\frac{1}{c(\beta)}},\hspace{3mm}\text{if $n$ is even}
	\end{array}
	\right.
	\end{split}\\[6mm]
&=&\begin{split}
	\left\{
	\begin{array}{lr}
\exp[\frac{\sum_{k=1}^{\frac{n-1}{2}}C_n^k}{c(\beta)}
(\ln(\frac{k^\alpha}{n^\alpha}+\frac{(n-k)^\alpha}{n^\alpha}-1))],\hspace{3mm} \text{if $n$ is odd}\\[1mm]
\exp[\frac{\sum_{k=1}^{\frac{n}{2}-1}C_n^k}{c(\beta)}
\ln(\frac{k^\alpha}{n^\alpha}+\frac{(n-k)^\alpha}{n^\alpha}-1)
+\frac{C_n^{\frac{n}{2}}\ln(2^{1-\alpha}-1)}{2c(\beta)}],\hspace{3mm}\text{if $n$ is even.}
	\end{array}
	\right.
	\end{split}
\end{array}
$$
\end{widetext}

\begin{figure}[t]
\centering
\includegraphics[width=8cm]{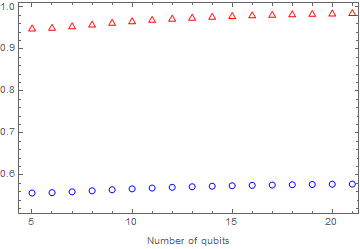}\\
\caption{The $\mathcal{G}_{1/3}(\ket{W_n})$ is given by the blue circles, and $\frac{\mathcal{G}_{1/3}(\ket{W_n})}{\mathcal{G}_{1/3}(\ket{GHZ_n})}$ is given by the red triangles for $n$ from $5$ to 21.}
\label{fig1}
\end{figure}

By using the Stolz-Cres\`{a}ro theorem, we have
$$\lim_{k\rightarrow\infty}\frac{\mathcal{G}_\alpha(\ket{W_{2k}})}
{\mathcal{G}_\alpha(\ket{GHZ_{2k}})}=
\lim_{k\rightarrow\infty}\frac{\mathcal{G}_\alpha(\ket{W_{2k+1}})}
{\mathcal{G}_\alpha(\ket{GHZ_{2k+1}})}=1.
$$
From Fig.\ref{fig1} we see that ${\mathcal{G}_{1/3}(\ket{W_n})}/{\mathcal{G}_{1/3}(\ket{GHZ_n})}$ is less than $1$ and tends to $1$ with the increase of $n$. It implies that the state $\ket{GHZ_n}$ is more genuine multipartite entangled than $\ket{W_n}$ in terms of G$\alpha$C, which is in consistent with the results of Ref.\cite{PRL112160401,vrana2015asymptotic} and shows that the measure G$\alpha$C captures the essence of GME for each $\alpha$.

\section{Entanglement ordering with G$\alpha$C}\label{s3}
As a genuine multipartite entanglement measure, G$\alpha$C gives a definite entanglement ordering for any pair of multipartite quantum states that can transform from one to the other with deterministic LOCC. Nevertheless, different measures may give rise to different ordering for even states in the same SLOCC equivalent class \cite{PRL83436,virmani2001optimal,zyczkowski2002relativity}. Let $E_1$ and $E_2$ be two measures of entanglement. For a pair of states $\sigma_1$ and $\sigma_2$, it is possible that $E_1(\sigma_1)\ge E_1(\sigma_2)$, while $E_2(\sigma_1)\le E_2(\sigma_2)$. Namely, $E_1$ and $E_2$ lead to different entanglement order. We study next the entanglement ordering given by
G$\alpha$C and some other GME measures by detailed examples, and show that G$\alpha$C possesses advantages over other GME measures in characterizing GME for certain types of entanglement.

$\mathit{Example\ 2}$.
Consider the following two types of three-qubit states:
\begin{align*}
\text{Type A:}~\ket{\psi}=&\frac{1}{\sqrt{2}}(\cos\theta\ket{000}+\sin\theta\ket{001})+\frac{1}{\sqrt{2}}\ket{111},\\
\text{Type B:}~\ket{\phi}=&\cos\theta\ket{000}+\sin\theta\ket{111},
\end{align*}
where $\theta\in[0,\frac{\pi}{2}]$.

\begin{figure}
\centering
\includegraphics[width=85mm]{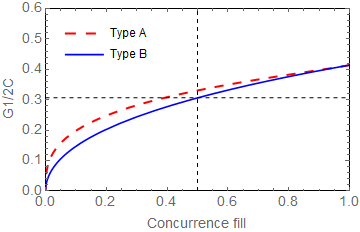}\\
\caption{The G$\alpha$C ($\alpha=1/2$) versus Concurrence fill for the type-A states and type-B states, respectively. Each point in a curve represents a three-qubit pure state. The dashed (red) denotes the type-A states, the solid (blue) stands for the type-B states.}
\label{f2}
\end{figure}

\begin{figure}
\centering
\includegraphics[width=85mm]{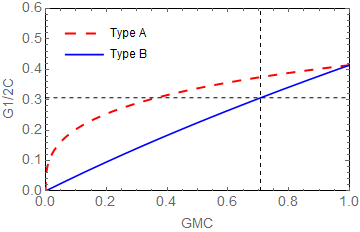}\\
\caption{The G$\alpha$C ($\alpha=1/2$) versus GMC for the type-A states and type-B states, respectively. Each point in a curve represents a three-qubit pure state. The dashed (red) denotes the type-A states, the solid (blue) stands for the type-B states.}
\label{f3}
\end{figure}

As shown in Figs. \ref{f2} and \ref{f3}, there are pairs of type-A and type-B states that are indistinguishable by the either the concurrence fill or the GMC, but can be distinguished by G$\alpha$C ($\alpha=1/2$), and vice versa. Namely, a state of type-A and a state of type-B may have the same value of concurrence fill or GMC, but have different values of G$\alpha$C ($\alpha=1/2$).

\begin{figure}
\centering
\includegraphics[width=85mm]{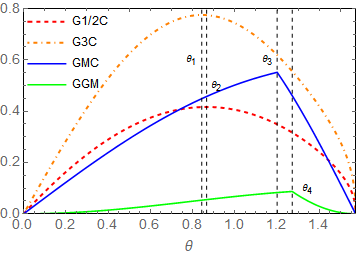}\\
\caption{The G$\frac{1}{2}$C, G$3$C, GMC and GGM for the four qubit pure states given in (\ref{e12}). The dashed (red) denotes the G$\frac{1}{2}$C. The dotdashed (orange) is for G$3$C. The solid (blue) stands for the GMC, and the solid (green) indicates the GGM.}
\label{f4}
\end{figure}

$\mathit{Example\ 3}$.
Consider the following family of four-qubit pure states,
\begin{align}\label{e12}
\ket{\varphi}=&\sin\theta(\cos\frac{3\pi}{5}\ket{0100}+\sin\frac{3\pi}{5}\ket{1000})\nonumber\\
&+\cos\theta\ket{0011}
\end{align}
for any $\theta\in[0,\frac{\pi}{2}]$.
From Fig. \ref{f4} we see that G$q$C ($q=3$), G$\alpha$C ($\alpha=1/2$), GMC and GGM reach their maximums at $\theta_1=0.843$, $\theta_2=0.866$, $\theta_3=1.199$, and $\theta_4=1.271$, respectively. As $\theta$ increases from $\theta_1$ to $\theta_2$, the G$3$C decreases, while the G$\frac{1}{2}$C increases, giving rise to opposite entanglement ordering in this region. As $\theta$ increases from $\theta_2$ to $\theta_3$ ($\theta_4$) the G$\frac{1}{2}$C decreases while the GMC (GGM) increases. Thus for two arbitrary states in this range, the entanglement ordering of G$\frac{1}{2}$C is the opposite to GMC (GGM).
There are pairs of states with the same G$3$C, GMC or GGM, but have different G$\frac{1}{2}$C, meaning that G$\frac{1}{2}$C is able to detect the difference of GME while G$3$C, GMC or GGM fail to. Moreover, sharp peaks of GMC and GGM emerge with the varying $\theta$, respectively, while the G$\frac{1}{2}$C and G$3$C are smooth with a continuous slope.

\smallskip
\section{Conclusion}\label{se4}
We have introduced G$\alpha$C, a class of well-defined GME measures with one parameter $\alpha$ for arbitrary multipartite states. By utilizing the symmetry in the definition of G$\alpha$C, analytical results have been derived for $n$-qubit GHZ states and W states. Based on G$\alpha$C, it is shown that the GHZ states are more genuinely entangled than the W states.
By explicit examples, we have demonstrated that G$\alpha$C can distinguish different GME states that other GME measures cannot. Hence, the G$\alpha$C serves as a complement to the current GME measures and enriches the theory of quantum multipartite entanglement. Our results may highlight further researches on the study of quantifying quantum multipartite entanglement and the physical understanding of quantum multipartite correlations. It would be also interesting to find ways of experimental estimation of G$\alpha$C or its lower bounds, as well as more potential applications of G$\alpha$C in quantum tasks involving GME.

\smallskip
\section*{Acknowledgments}
This work is supported by the National Natural Science Foundation of China (NSFC) under Grants 12075159 and 12171044, and the specific research fund of the Innovation Platform for Academicians of Hainan Province.

\end{document}